\documentclass[preprint]{aastex}
\usepackage{psfig}

\slugcomment{To appear in \it The Astrophysical Journal, Letters}

\shorttitle{Ionizing Photons in Clusters}
\shortauthors{Maloney \& Bland-Hawthorn}

\begin{document}

\def\ie{{\it i.e.,\ }}
\def\eg{{\it e.g.,\ }}
\def\qv{{\it q.v.,\ }}
\def\cf{{\it cf.\ }}
\def\etal{{\it et al.}}
\def\gtrapprox{\;\lower 0.5ex\hbox{$\buildrel >
    \over \sim\ $}}             
\def\lessapprox{\;\lower 0.5ex\hbox{$\buildrel < \over \sim\ $}}
\def\msol{\ifmmode {\>M_\odot}\else {$M_\odot$}\fi}
\def\Omegab{\ifmmode {\Omega_{\rm baryon}}\else {$\Omega_{\rm baryon}$}\fi}
\def\pyr{\ifmmode {\>{\rm\ yr}^{-1}}\else {yr$^{-1}$}\fi}
\def\psec{\ifmmode {\>{\rm\ s}^{-1}}\else {s$^{-1}$}\fi}
\def\kms{\ifmmode {\>{\rm km\ s}^{-1}}\else {km s$^{-1}$}\fi}
\def\psqcm{\ifmmode {\>{\rm cm}^{-2}}\else {cm$^{-2}$}\fi}
\def\emunits{\ifmmode {\>{\rm cm^{-6}\;pc}}\else {cm$^{-6}$ pc}\fi}
\def\pkunits{\ifmmode {\>{\rm cm^{-3}\;K}}\else {cm$^{-3}$ K}\fi}
\def\pcubcm{\ifmmode {\>{\rm cm}^{-3}}\else {cm$^{-3}$}\fi}
\def\hubunits{\ifmmode {\>{\rm km\ s^{-1}\ Mpc^{-1}}}\else {km
s$^{-1}$ Mpc$^{-1}$}\fi}
\def\phoflux{\ifmmode{{\rm phot\ cm}^{-2}{\rm\ s}^{-1}}\else {phot
cm$^{-2}$ s$^{-1}$}\fi}
\def\raylunit{\ifmmode{{\rm phot\ cm}^{-2}{\rm\ s}^{-1}{\rm\ sr}^{-1}}
\else {phot cm$^{-2}$ s$^{-1}$ sr$^{-1}$}\fi}
\def\phoemis{\ifmmode{{\rm phot\ cm}^3{\rm\ s}^{-1}{\rm sr}^{-1}}
\else{phot cm$^3$ s$^{-1}$ sr$^{-1}$}\fi}
\def\emisunit{\ifmmode{{\rm cm}^{-3}{\rm\ s}^{-1}{\rm sr}^{-1}}
\else{cm$^{-3}$ s$^{-1}$ sr$^{-1}$}\fi}
\newcommand{\hikpc}{{\hbox {$h^{-1}$}{\rm kpc}} }
\newcommand{\himpc}{{\hbox {$h^{-1}$}{\rm Mpc}} }
\newcommand{\0}{ {\scriptscriptstyle {0}} }
\newcommand{\T}{ {\scriptscriptstyle {\rm T}} }
\newcommand{\qrms}{ Q_{\scriptscriptstyle {\rm RMS}} }
\newcommand{\kev}{ {\rm keV} }
\newcommand{\mpc}{ {\rm Mpc} }
\newcommand{\dttsz}{{\hbox
      {$\displaystyle\left({\delta T \over T}\right)_\sz$} }}
\newcommand{\dtt}{{\hbox
      {$\displaystyle\left({\delta T \over T}\right)$} }}
\def\Tkev{\ifmmode{T_{\rm kev}}\else {$T_{\rm keV}$}\fi}
\def\Em{\ifmmode{{\cal E}_m}\else{${\cal E}_m$}\fi}
\def\Dm{\ifmmode{{\cal D}_m}\else{${\cal D}_m$}\fi}
\def\Ha{H$\alpha$}
\def\euve{{\it EUVE}}
\def\thetamin{\ifmmode{\theta_{\rm min}}\else {$\theta_{\rm min}$}\fi}
\def\thetamax{\ifmmode{\theta_{\rm max}}\else {$\theta_{\rm max}$}\fi}
\def\be{\begin{equation}}
\def\ee{\end{equation}}
\def\bea{\begin{eqnarray}}
\def\eea{\end{eqnarray}}


\title{Ionizing Photons and EUV Excesses in Clusters of Galaxies}


\author{Philip R. Maloney\altaffilmark{1}}
\affil{Center for Astrophysics \& Space Astronomy, 
University of Colorado,
Boulder, Colorado, 80309-0389}
\altaffiltext{1}{maloney@casa.colorado.edu}
\and
\author{J. Bland-Hawthorn\altaffilmark{2}}
\affil{Anglo-Australian Observatory, P.O. Box 296,
Epping, NSW 2121, Australia}
\altaffiltext{2}{jbh@aaoepp2.aao.gov.au}

\begin{abstract}
Observations with the {\it Extreme Ultraviolet Explorer} satellite
are purported to show extreme ultraviolet (EUV) and soft X-ray
excesses in several clusters of galaxies (Bonamente, Lieu \& Mittaz
2001). If interpreted as thermal emission, this would imply the
presence of warm ($T\sim 10^6$ K) gas in these clusters with a mass
comparable to that of gas at coronal temperatures. If true, this would
have profound implications for our understanding of galaxy clusters
and the distribution of baryons in the universe. Here we show that
because of the large ionizing photon emissivities of gas at such low
temperatures, the ionizing photon fluxes seen by disk galaxies in the
observed clusters can be very large, resulting in minimum emission
measures from neutral gas in such disks as high as $100\emunits$. This
result is essentially independent of the mechanism actually
responsible for producing the alleged EUV excesses. The predicted
emission measures in Abell 1795 ($z=0.063$) are about an order of
magnitude larger than seen in the Reynolds layer of the Galaxy,
providing a straightforward observational test of the reality of the
EUV excess. New tunable filter \Ha\ images and WFPC images from the
Hubble Space Telescope archive do not support the existence of the
claimed EUV excess.
 
\end{abstract}


\section{Introduction}
Rich clusters of galaxies are one of the major identified reservoirs
of baryons at low redshift; the contribution of hot plasma in clusters
to \Omegab\ is comparable to that of the stellar components of
galaxies (Fukugita, Hogan \& Peebles 1998). Most of the intracluster
medium (ICM) in galaxy clusters is at temperatures $T\sim 1-10$ keV, as
expected from the depth of the potential wells traced by the
galaxies, although the high-density cores of clusters may exhibit gas
at lower temperatures (\eg Sarazin 1988). This hot gas radiates
primarily through thermal bremsstrahlung. Since bremsstrahlung is
intrinsically broad-band emission, the ICM should be detectable at
soft X-ray and extreme ultraviolet (EUV) wavelengths as well as keV
energies. 

The {\it Extreme Ultraviolet Explorer} (\euve) satellite provided the
capability to observe at energies of order 100 eV; the precise
bandpass depends on convolution of the instrument response with the
absorption due to the intervening interstellar medium.  \euve\
observations of several clusters of galaxies have been interpreted as
implying substantial EUV emission in excess of that expected from the
gas observed at keV energies (Lieu \etal\ 1996a,b; Mittaz, Lieu \&
Lockman 1998). This claim has been disputed by Bowyer, Bergh\"ofer \&
Korpela (1999) who argue that the EUV excesses are an artifact caused
by improper subtraction of the instrumental background (although they
infer that the relatively weak EUV excesses in the Virgo and Coma
clusters may be real), and an increasingly fractious debate has ensued
(\eg Lieu \etal\ 1999; Bonamente, Lieu \& Mittaz 2001, hereafter BLM;
Bergh\"ofer, Bowyer, \& Korpela 2000; Bonamente \etal\ 2001). Dixon,
Hurwitz, \& Ferguson (1996) and Dixon \etal\ (2001) used the {\it
Hopkins Ultraviolet Telescope} (HUT) and {\it Far-Ultraviolet
Spectroscopic Explorer} (FUSE), respectively, to search for the far-UV
resonance lines of C IV $\lambda\lambda 1548$, 1551 and O VI
$\lambda\lambda 1032$, 1036 (both the carbon and oxygen lines in the
former paper and just the oxygen lines in the latter), which would be
expected to be prominent if the EUV excesses are produced by warm
gas. Neither observation detected any line emission; Dixon \etal, in
particular, ruled out the existing warm gas models for the Virgo and
Coma clusters at the $2\sigma$ level. In this {\it Letter}, we
present an alternative probe of the reality of the EUV excesses,
namely, observations of recombination line radiation produced by
irradiation of neutral gas in galaxies by the ionizing radiation
associated with the EUV excesses.


\section{Cluster Ionizing Photon Fluxes}
Two different mechanisms have been proposed to explain the EUV
excesses: thermal emission from warm ($T\sim 10^6$ K) gas (Lieu \etal\
1996a), and inverse Compton (IC) scattering of cosmic microwave
background photons by nonthermal electrons in the cluster (Hwang 1997;
Ensslin \& Biermann 1998). Both of these mechanisms suffer from
serious difficulties. In the thermal model, the gas cooling times are
so short that it is difficult to see how such emission could be
sustained for more than a small fraction of the age of the
cluster. Fabian (1997) proposed that the EUV emission arises in mixing
layers, so that the energy source for the emission is the thermal
reservoir of hot gas. This could alleviate the cooling difficulty for
several of the clusters but probably fails for the most EUV-luminous
cluster, Abell 1795 (A1795). The nonthermal model sidesteps this
problem; however, it runs into difficulties with the strength of
cluster magnetic fields (\eg Kempner \& Sarazin 2000) and the pressure
of nonthermal particles (including the relativistic electrons) (Bowyer
\& Bergh\"ofer 1998; BLM).

We adopt the thermal interpretation of the EUV excesses since BLM have
presented detailed thermal fits (gas density and temperature) as a
function of radius to the EUV emission from the Virgo and A1795
clusters. As these fits by definition reproduce the alleged EUV
excesses in these clusters, the resulting ionizing photon fluxes
$\phi_i$ should not be very sensitive to this choice as opposed to an
IC model. (Since the emission from a power-law electron distribution
will have a broader energy distribution than thermal emission, which
is dominated by line emission at these temperatures, a thermal model
will produce smaller values of $\phi_i$ than an IC model.)

For simplicity, we assume a spherically symmetric distribution of gas;
because of the integration over angle and radius, we expect the
results to be insensitive to this assumption. Specifically, we take
the radial distribution of protons to be described by a
$\beta-$model, 
\be
n(r)=n_o\left[1+(r/r_o)^2\right]^{-3\beta/2}
\ee
where $n_o$ and $r_o$ are the core density and radius, respectively.
Assuming thermal emission from the
gas, the ionizing photon emissivity is proportional to $n^2$, and the
normally incident ionizing photon flux on a surface at radius $R$
oriented normally to the radius vector is given by
\be
\phi_i=2\pi \xi_i n_o^2 r_o x_o^{6\beta-1}
\int_\thetamin^\thetamax\int_0^{x_m}d\theta\,dx\,\sin\theta\cos\theta\,
A_x^{-3\beta}\qquad \phoflux
\ee
where $x=r/R$, $x_o=r_o/R$, and $A_x\equiv 1+x_o^2+x^2-2x\cos\theta$. 
The ionizing photon emissivity $\xi_i$ is defined so that $n^2\xi_i$
is the number of ionizing photons emitted \emisunit. The integral over
the radial variable can be done analytically for $3\beta$ equal to an
integer or half-integer. The limits of the angular integration depend
on the hemisphere of integration; $(\thetamin,\thetamax)=(0,\pi/2)$
gives the flux on the inward-facing surface, while $(\pi/2,\pi)$ gives
the flux on the outward surface. The maximum radius of integration
$x_m$ is $\theta-$dependent for a finite gas distribution.

For consistency with the thermal fits in BLM, we have used the
the MEKAL plasma code (Kaastra, Mewe, \& Nieuwenhuijzen 1996, as
implemented in the SPEX software package: Kaastra \etal\ 1995) to
calculate the ionizing photon emissivities for the cluster gas. In
figure 1 we plot $\xi_i$ as a function of the gas temperature; we have
taken the upper limit to be 500 eV in calculating $\xi_i$, but the
results are insensitive to this assumption. Metal abundances of $1/3$
solar have been assumed. For comparison, we have also plotted $\xi_i$
as calculated using the MAPPINGS III code (Sutherland \& Dopita 1993;
Dopita \& Sutherland 1996). The largest differences are about $40\%$,
which we take to be a measure of the uncertainty in the predicted
$\phi_i$.

In Figure 2 we plot the normally incident photon flux $\phi_i$ on a
surface as a function of radius for the A1795 cluster. (This is the
sum of the fluxes incident on opposite sides of the surface.) The
upper solid curve uses the thermal fit to the EUV emission from
BLM. Because this result is sensitive to the density at the largest
radius for which BLM present results, we have also plotted $\phi_i$
for this model with the gas density in the outermost bin reduced
by its stated uncertainty (dashed line). Finally, the lower solid
curve shows the ionizing photon flux expected from the hot
X-ray-emitting gas in the cluster, with the $T$ and
$n$ as derived from XMM-Newton observations (Tamura \etal\
2001)\footnote{Since we actually truncate the cluster gas distribution
at $10'$ radius, the limit of the fits presented by BLM, the curves in
Figure 2 are lower limits to the true ionizing fluxes; this effect is
much more important for the EUV excess models.}. Although, as noted
above, the curves are for a surface oriented normally to the radius
vector, the extremely weak dependence on radius for the models with
EUV excesses indicates that the actual fluxes incident on galactic
disks at random orientations within the cluster will not differ
significantly from the plotted values. 

\section{Discussion}
From the plot of ionizing photon fluxes in Figure 2, we see that
$\phi_i$ will be a minimum of one to two orders of magnitude larger if
the EUV excess claimed in A1795 by BLM is present than it would be if
there is only the more conventional hot gas component; over most of
the cluster volume, the ratio is actually much larger, $\sim
10^2-10^3$, since $\phi_i$ from the EUV component is much less
centrally concentrated than the hot gas. Any spiral galaxy within the
inner $10'$ ($660h_{70}^{-1}$ kpc at the distance of
A1795) will see an ionizing photon flux $\phi\approx 5\times 10^7$
\phoflux\ (for the warm gas parameters derived by BLM). The
corresponding emission measure (normal to the disk )
$\Em\approx 63\emunits$, assuming $T_e\approx 10^4$ K (\eg Maloney \&
Bland-Hawthorn 1999). For comparison, this is about ten
times larger than the emission measure characterizing the warm ionized
medium (the Reynolds layer: Reynolds 1984).

This offers a straightforward observational test of the reality of the
EUV excesses. The large predicted emission measures can be easily
probed through recombination line observations. In particular, the
predicted \Ha\ surface brightness is \be I_{\rm H\alpha}=35.7
\left({\Em\over 100\emunits}\right)\;{\rm Rayleigh} \ee where
$1\;{\rm Rayleigh}=10^6/4\pi$ \raylunit, and we have
normalized to the expected \Em\ for galaxies in
A1795\footnote{We have performed similar calculations for the Virgo
cluster, for which BLM also present detailed fits, but we do not
present them as the results are uninteresting: the EUV excess
contribution does not exceed that from the hot gas in the cluster
(which has $\Tkev\sim 1$), and the ionizing photon flux drops below
the cosmic background level at a distance smaller than the separation
of any disk galaxy from the center of the emission. This is largely a
consequence of the small spatial scale of the emission, which is
concentrated on the giant elliptical galaxy M87.}. 
The resulting $I_{\rm H\alpha}$ depends
only on $\phi_i$ and is independent of $T_e$ and $n$. Emission measures
as low as unity can be reached quite readily with a differential
narrow-band imager, in particular, the TAURUS Tunable
Filter\footnote{See the TTF web site at http://www.aao.gov.au/ttf/}
(TTF) which combines beam switching and frequency switching to suppress
atmospheric variability (Bland-Hawthorn \& Jones 1998). These faint
emission levels can also be reached in one night with
spectrographs on 8m class telescopes (Cirkovic, Bland-Hawthorn, \&
Samurovic 1999).

The cluster-induced \Ha\ emission can be easily distinguished from the
warm ionized media intrinsic to the cluster galaxies, even if we adopt
a reduced density in the outer regions of the cluster (dashed curve in
Figure 2). The surface brightness of the diffuse ionized gas in spiral
disks are typically of the same order as in the Reynolds layer, with
$\Em\sim 10$ \emunits\ (\eg Ferguson \etal\ 1996; Hoopes, Walterbos,
\& Greenawalt 1996; Greenawalt, Walterbos, \& Braun 1997; Greenawalt
\etal\ 1998), and decline with increasing radius, generally to
$\Em\sim 1$ \emunits\ around the edge of the optical disk. The radial
behavior of the cluster-induced \Ha\ will be very different. Since the
cluster ionizing field will be uniform across a galactic disk, the
entire neutral hydrogen disk of a galaxy, even outside the optical
disk, should be glowing in \Ha\ at the same level. Hence this
component will be flat with radius, and extend outside of the optical
disk, as far as the neutral hydrogen edge. We also note that even such
high ionizing photon fluxes probably do not significantly increase the
critical column density for ionization of galactic neutral hydrogen
disks above the value expected in field spirals due to the cosmic
ionizing background (Maloney 1993), since the pressure of the ICM will
force the ISM of cluster spirals to high values ($P/k\gtrapprox 10^4$
\pkunits) even at large radius.

This technique would fail if stripping of gas from spiral galaxies in
the cluster were a very efficient process, as suggested by early HI
observations of the densest clusters (Giovanelli \& Haynes 1985;
Cayette \etal\ 1990). However, more recent studies show that stripping
is less effective than originally suggested. The most comprehensive
survey to date, incorporating 1900 galaxies in 18 clusters (Solanes
\etal\ 2001), finds that more than one-half of spirals retain at least
half their HI within $\sim 900 h_{70}^{-1}$ kpc ($13'$ at the distance
of A1795). This is the scale, roughly $40\%$ of an Abell radius, over
which we expect to see disks with bright H-alpha disks or halos (see
Fig. 2). A third of the clusters in the Solanes \etal\ sample show no
H$\;$I deficiency.

We have investigated the brighter end of the predicted range of
$\phi_i$ using archival broad-band images of the core of A1795. An
emission measure of \Em\ produces an R band surface brightness of
$\mu_R \approx 31.5 - 2.5\log \Em$ mag arcsec$^{-2}$. An extended HI
disk seen face-on glowing at $\Em \approx 100$ cm$^{-6}$ pc
corresponds to $\mu_R \approx 26.5$; this rises to $\mu_R \approx 25$
for a disk viewed edge on. We have reanalyzed the {\it Hubble Space
Telescope} (HST) WFPC2 images of A1795 obtained by J.~Trauger (ID
5212) (\qv McNamara \etal\ 1996). The flux-calibrated F702W images
have a summed exposure time of 1780 sec. The combined image achieves
$\mu_R \approx 25$ mag arcsec$^{-2}$ at S/N $\approx 0.5$ per WFPC
pixel. When smoothed over $1''$, we see no evidence of faint, flat
\Ha\ halos down to $\mu_R \approx 26.5$ ($2\sigma$). The disk galaxies
appear to be exponential down to and below this level. (The HST has
been used to reach down to 28 mag arcsec$^{-2}$ in broadband images,
\eg Tyson \etal\ 1998. At these levels, it would be possible to detect
\Ha\ halos with $\Em \approx 10$ cm$^{-6}$ pc in inclined galaxies.)
However, the WFPC2 L-shaped field of view is only about $3'$.

We have attempted a second experiment with the TTF at the
Anglo-Australian Telescope (AAT) in order to reach fainter levels in
\Ha\ over a $10'$ diameter field. A1795 was observed for us by A.~Edge
and K.~Baker on 2001 Mar 3 using the TTF at the AAT f/8 Cassegrain
focus. The tunable filter was set to 14\AA\ effective bandpass.
To suppress systematic errors induced by the atmosphere, we
used the TTF in `straddle shuffle' mode. The on band
exposures (6970\AA) were interleaved with blue and red off band
exposures (6920\AA,7050\AA). 
This differential imaging mode produces
perfect off-band subtraction (zero systematic error) not possible with
conventional imaging filters (Bland-Hawthorn \& Jones 1998).
Two observing sequences gave a total exposure time of 20
minutes each on the redshifted \Ha\ line and 
the summed off band (Figure 3).

While the photometric conditions were good, the seeing was variable
around $1.6''$. The flux-calibrated data reach a surface brightness
$E_m \approx 5$ \emunits\ at 2$\sigma$. {\it At these faint levels we
find no evidence of diffuse \Ha\ disks or extended \Ha\ halos for
galaxies over the $10'$ field.} There is evidence for nucleated star
formation in some galaxies, and the central dominant elliptical shows
evidence of a spectacular cooling flow (Edge \etal\ 2001, in
preparation).

The preliminary observations from HST and the TTF do
not support the existence of faint extended halos or disks within $5'$
($350 h_{70}^{-1}$ kpc) of the cluster core. 

\section{Summary}
We present an independent method for determining the reality of the
alleged EUV excesses in galaxy clusters. At least for the case of
Abell 1795, the ionizing photon fluxes implied by the claimed EUV
emission imply that the neutral hydrogen disks of spiral galaxies
within the cluster (out to at least $10'$ radius from the cluster
center, corresponding to about 660 kpc) should be glowing in hydrogen
recombination lines, with surface brightnesses of tens of
Rayleighs at \Ha. Current instrumental sensitivities should make
it possible to detect the \Ha\ emission expected due to the
ionizing photon flux from the hot gas in the cluster (which is two
orders of magnitude or more weaker than the predicted fluxes in the
EUV excess model) out to several arcminutes radius, and hence
\Ha\ observations can easily discriminate between the two cases
and provide stringent limits on the presence of any additional source
of ionization for spiral galaxies within the cluster. A preliminary
investigation using archival broad-band images and new \Ha\ 
observations does not support the existence of the alleged EUV excess
in this cluster.

We are very grateful to Alastair Edge and Kurt Baker for allowing us
to use their H$\alpha$ data in advance of publication and to the
referee for a helpful and extraordinarily prompt report. JBH would
like to thank Roberto de Propris, John Dickey, Greg Taylor, and
Warrick Couch for useful conversations. PRM is supported by the
National Science Foundation under grant AST 99-00871.

\clearpage
\setcounter{figure}{0}
\begin{figure}
\centerline{\psfig{figure=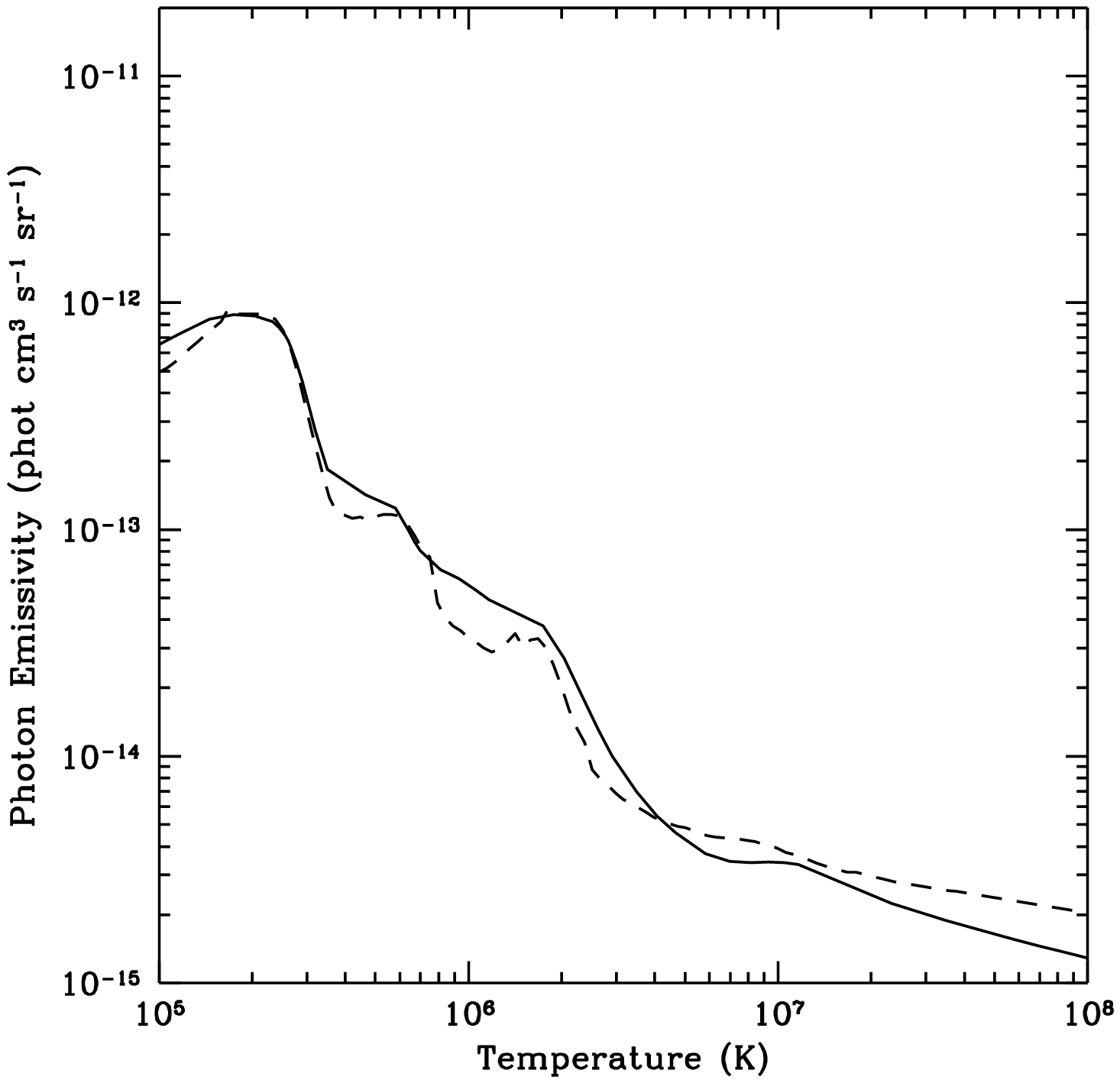,width=\textwidth,angle=0}}
\caption{Ionizing ($13.6-500$ eV) photon 
emissivity $\xi_i$ (\phoemis) as a function of gas temperature, as
calculated with MEKAL (solid line) and MAPPINGS III (dashed line).}
\end{figure}

\clearpage
\begin{figure}
\centerline{\psfig{figure=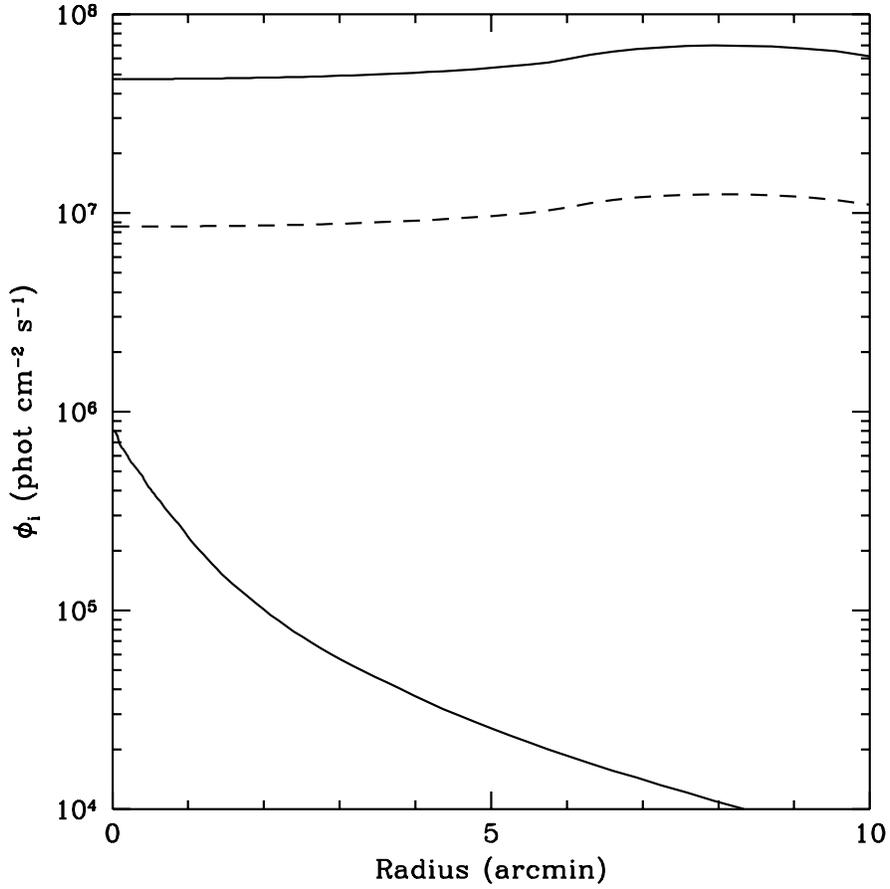,width=\textwidth,angle=0}}
\caption{Normally incident ionizing photon fluxes $\phi_i$ as a
function of radius in the A1795 cluster, as derived from thermal fits
to the EUV excess claimed by BLM (upper solid line) and from fits to
the X-ray emission observed with XMM-Newton (Tamura \etal\ 2001) (lower
line). The dashed curve shows $\phi_i$ for the EUV excess fit with the
density in the outermost bin reduced by its stated uncertainty. Only
the contribution from gas within $10'$ of the cluster center has been
included.}
\end{figure}

\clearpage
\begin{figure}
\centerline{\psfig{figure=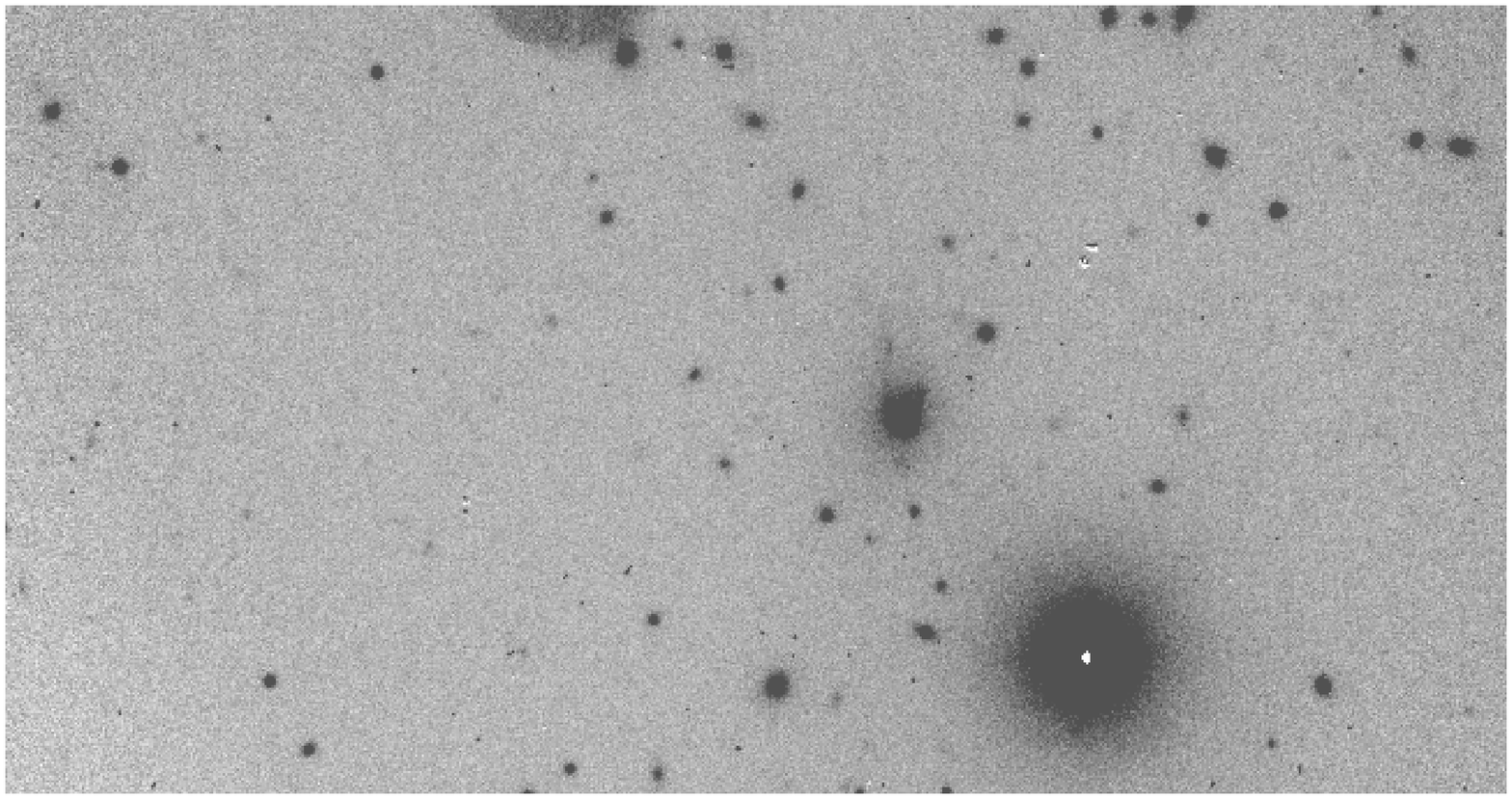,width=\textwidth,angle=0}}
\centerline{\psfig{figure=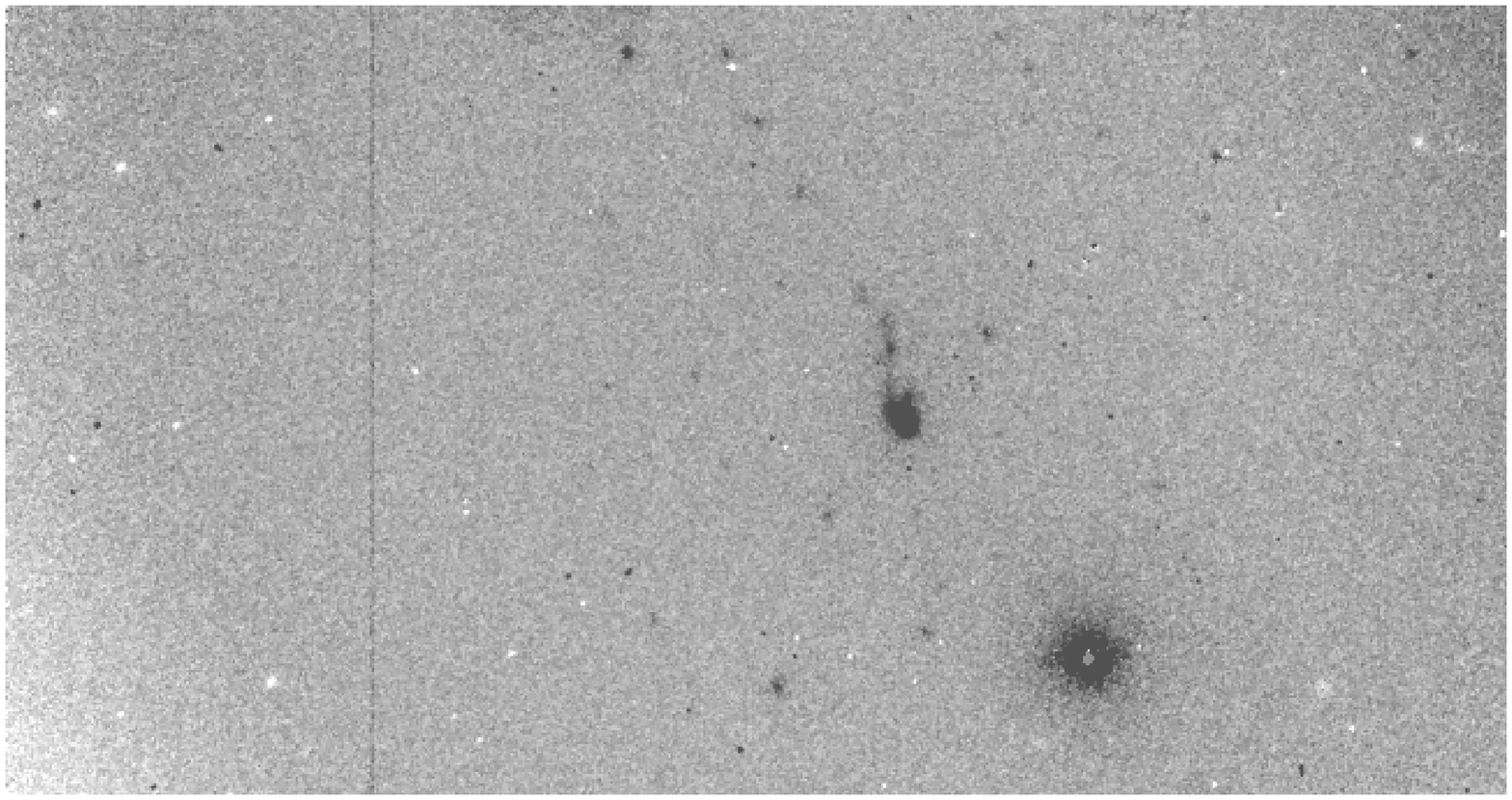,width=\textwidth,angle=0}}
\caption{Sections of the TTF `straddle shuffle' image of A1795 showing
H$\alpha$ emission at the cluster redshift ($z=0.063$). The upper
strip is the sum of the on and off bands. (The half annulus is an
out-of-focus ghost of the bright star to the SW.)  The lower strip is
the continuum-subtracted line band.  The field of view for each strip
is 4.2' by 8.0'. Galaxies show either weak Balmer absorption or
nucleated H$\alpha$ emission. The central giant elliptical has a
spectacular cooling flow with H$\alpha$ filaments characterized by
$E_m = 5-100$ cm$^{-6}$ pc.}
\end{figure}

\end{document}